\documentclass[preprint]{elsarticle}

\usepackage{lineno,hyperref}
\modulolinenumbers[5]
\usepackage{dcolumn}
\usepackage{graphicx}
\usepackage{color}
\usepackage{multirow}
\usepackage{textcomp}
\usepackage{physics}
\usepackage{dsfont}
\usepackage{epstopdf}
\usepackage{amsmath}

\journal{Phys. Lett. B}
\begin{document}
	\newcommand {\nc} {\newcommand}
\nc {\beq} {\begin{eqnarray}}
\nc {\eeq} {\nonumber \end{eqnarray}}
\nc {\eeqn}[1] {\label {#1} \end{eqnarray}}
\nc {\eol} {\nonumber \\}
\nc {\eoln}[1] {\label {#1} \\}
\nc {\ve} [1] {\mbox{\boldmath $#1$}}
\nc {\ves} [1] {\mbox{\boldmath ${\scriptstyle #1}$}}
\nc {\mrm} [1] {\mathrm{#1}}
\nc {\half} {\mbox{$\frac{1}{2}$}}
\nc {\thal} {\mbox{$\frac{3}{2}$}}
\nc {\fial} {\mbox{$\frac{5}{2}$}}
\nc {\la} {\mbox{$\langle$}}
\nc {\ra} {\mbox{$\rangle$}}
\nc {\etal} {\emph{et al.}}
\nc {\eq} [1] {(\ref{#1})}
\nc {\Eq} [1] {Eq.~(\ref{#1})}
\nc {\Refc} [2] {Refs.~\cite[#1]{#2}}
\nc {\Sec} [1] {Sec.~\ref{#1}}
\nc {\chap} [1] {Chapter~\ref{#1}}
\nc {\anx} [1] {Appendix~\ref{#1}}
\nc {\tbl} [1] {Table~\ref{#1}}
\nc {\Fig} [1] {Fig.~\ref{#1}}
\nc {\ex} [1] {$^{#1}$}
\nc {\Sch} {Schr\"odinger }
\nc {\flim} [2] {\mathop{\longrightarrow}\limits_{{#1}\rightarrow{#2}}}
\nc {\textdegr}{$^{\circ}$}
\nc {\inred} [1]{\textcolor{red}{#1}}
\nc {\inblue} [1]{\textcolor{blue}{#1}}
\nc {\IR} [1]{\textcolor{red}{#1}}
\nc {\IB} [1]{\textcolor{blue}{#1}}
\nc{\pderiv}[2]{\cfrac{\partial #1}{\partial #2}}
\nc{\deriv}[2]{\cfrac{d#1}{d#2}}

\begin{frontmatter}

\title{Sensitivity of one-neutron knockout observables of loosely- to more deeply-bound nuclei}

\author[FRIB]{C.~Hebborn}
\cortext[mycorrespondingauthor]{Corresponding author}
\address[FRIB]{Facility for Rare Isotope Beams, Michigan State University, East Lansing, MI 48824, USA}
\ead{hebborn@frib.msu.edu}

\author[JGU]{P.~Capel}
\address[JGU]{Institut f\"ur Kernphysik, Johannes Gutenberg-Universit\"at Mainz, D-55099 Mainz}
\ead{pcapel@uni-mainz.de}

\begin{abstract}
For the last few decades, one-nucleon knockout reactions on {light} composite targets---$^9$Be or $^{12}$C---have been extensively  used to study  the single-particle (s.p.) structure of nuclei far from stability. To determine which information can be accurately inferred  from knockout cross sections, we conduct a  sensitivity analysis of these  observables considering various s.p.\ descriptions {within the usual eikonal description of the reaction}.
This work  shows  that {total} one-neutron knockout cross sections are not sensitive to the short-range part of the s.p.\ wave function.
{Rather, they} scale with the {mean square radius} of the overlap function.
Using a perturbative expression of the cross section, we can easily explain our numerical predictions {analytically}.
This analysis  suggests   that (i) spectroscopic factors extracted from knockout data {suffer} from sizeable model uncertainties associated with the choice of s.p.\ wave functions and (ii) knockout reactions constitute an excellent probe of the radius of the nucleus and therefore offer an alternative technique to infer the neutron-skin thickness of exotic nuclei.
\end{abstract}

\begin{keyword}
Knockout reactions, spectroscopic factors, nuclear radius, eikonal approximation
\end{keyword}

\end{frontmatter}


The structure of nuclei away from stability challenges the usual description of nucleons piling up in well-defined shells to form a compact object \cite{Oetal20}.  Due to their short lifetime, unstable nuclei are often studied through reactions, i.e., they are synthetized and sent onto a target before they decay. Structure information is  inferred from the comparison of experimental cross sections to theoretical predictions.
In this Letter, we focus on one-nucleon knockout (KO).
This reaction corresponds to the removal of one nucleon from the projectile through its interaction with a light target nucleus (typically $^9$Be or $^{12}$C) at intermediate beam energy, viz.\ 60--100~MeV/nucleon.
{Because usual models of the reaction express the cross sections as a direct function of the single-particle (s.p.)} state of the removed nucleon, KO has been extensively used to study the s.p.\ structure of unstable nuclei \cite{HS01,HT03,GG08}.

The analysis of knockout measurements typically relies on the core-spectator approximation \cite{HT03,HM85}, which assumes that, after the one-neutron removal, the core of the nucleus is in the state it had within the projectile.
Measuring the final state of the core should thus give insights on the s.p.\ configuration of the projectile.
That configuration is characterised by the different orbitals $n_rlj$ within which the removed nucleon can be in the nucleus; here $l$ is the orbital angular momentum for the nucleon-core relative motion, $j$ is its total angular momentum obtained from the coupling of $l$ with the nucleon spin $s$, and $n_r$ is the number of nodes in the radial wave function.
Accordingly, the cross section for the removal of a nucleon from an initial state $i$ leaving the core in a final state $f$, is obtained as a linear {combination of products of spectroscopic factors $S^f_{n_rlj}$, usually obtained within the shell model, and s.p.\ knockout cross sections} $\sigma_{\rm KO}$ computed for a nucleon in orbital $n_rlj$ with a unit spectroscopic factor \cite{HT03}
\begin{equation}
\sigma_{\rm th}^f= \sum_{n_rlj}S^f_{n_rlj}\ \sigma_{\rm  KO}(n_rlj)\label{eq1}.
\end{equation}
The s.p.\ knockout cross {sections $\sigma_{\rm KO}$ are} given by the sum of  two contributions: the \emph{diffractive breakup} ($\sigma_{\rm bu}$) in which the target stays in its ground state, and the \emph{stripping} ($\sigma_{\rm str}$) which describes all the channels where the neutron is absorbed by the target.
These two contributions are usually obtained at the eikonal approximation, which takes as input effective projectile-target interactions and a s.p.\ wave function to describe the projectile \cite{HT03,HM85,G59}.

Although  one-nucleon knockout reactions have provided valuable insights on the evolution of the shell structure away from stability, their {total} cross sections are still not well understood.
In particular, the ratio between the experimental  cross sections and theoretical predictions, $R_s=\sigma_{\rm exp}/\sigma_{\rm th}$ displays a systematic linear dependence with the neutron-to-proton asymmetry of the nucleus {$\Delta S$}, which is not observed in other reactions, such as quasifree scattering $\rm (p,2p)$ and $\rm (p,pn)$ \cite{GM18} and transfer reactions~\cite{PhysRevLett.104.112701,PhysRevLett.129.152501,HNLPRL22}.
For the knockout of loosely-bound nucleons $R_s\sim 1$, whereas for more deeply-bound nucleons  $R_s\sim0.3$ \cite{Gade:08,TG21}.
This suggests a quenching of the spectroscopic factor for deeply-bound nucleons.
Because of this discrepancy, many studies have questioned the theoretical  description of  knockout reactions  \cite{Aetal21}. 
Some recent works argue that the  inaccuracy of the eikonal treatment of the reaction is responsible for the asymmetry dependence of $R_s$ \cite{HNLPRL22,Fetal12,Letal11,GOMEZRAMOS2022137252,HP2022}.
Other analyses have pointed out that part of this asymmetry dependence  might be due to the missing short- and long-range correlations in the shell-model description~\cite{PhysRevLett.103.202502,Jetal11,ATKINSON2019135027,Wetal21}.

To determine which nuclear-structure information can be accurately inferred from any reaction data, systematic studies using different overlap functions should be performed.
In Refs.~\cite{HC19,HC21ANC}, we have shown that knockout observables for loosely-bound one-neutron halo nuclei are peripheral, i.e., they are  sensitive only to the tail of the overlap function and not {just} to the square of its norm, namely its spectroscopic factor. 
In this work, we extend this study to the knockout of more deeply-bound neutrons. Our goal is to shed light on how  the peripherality of knockout observables evolves with the binding energy of the projectile. 
We analyse how those cross sections scale with the spectroscopic factor, the asymptotic normalisation constant (ANC) of the overlap wave function and its mean square radius.

We evaluate the s.p.\ knockout cross section using the usual few-body framework, in which the projectile $P$ is seen as  a structureless core $c$ and a valence neutron n, impinging on a target $T$, whose structure is also ignored~\cite{HT03,BC12}.  We consider two projectiles: first, a realistic one-neutron halo $^{11}$Be with a valence neutron bound by only 500~keV and, second, a fictional $^{11}$Be, in which the one-neutron separation energy is $S_{\rm n}=10$ MeV.
In both cases, we assume the $^{10}$Be core to be in its $0^+$ ground state with a spectroscopic factor 1.
The structure of these projectiles is modelled by an effective $c$-n Gaussian potential,  whose depth  is fitted to reproduce the desired separation energy of the projectile's  ground state.  The interactions between the projectile constituents and the target are simulated by optical potentials $U_{cT}$ and $U_{{\rm n}T}$, which include an imaginary term simulating all the inelastic channels not explicitly accounted for by the model.
We use here the same optical potentials as in Ref. \cite{HC19}. 

The stripping and diffractive-breakup contributions to the cross section are computed at the usual eikonal approximation \cite{HT03,HM85,G59,HBE96}.
The former reads 
	\begin{equation}
\sigma_{\rm str}=\frac{2\pi}{2j+1}\sum_{m}\int_0^\infty\mel{\phi_{n_rljm}}{(1-|S_{{\rm n}T}|^2)|S_{cT}|^2}{\phi_{n_rljm}} b\,db,\label{eq5}
\end{equation}
	where $\phi_{n_rljm}$ is the s.p.\ wave function generated by the Gaussian potential.
The sum is computed over $m$, the projection of the total angular momentum $j$.
The integral is performed over the impact parameter $b$, the transverse component of the  $P$-$T$ relative coordinate $\ve R$; the $Z$ direction is oriented along the incoming beam axis.
The diffractive-breakup cross section reads 
\begin{eqnarray}
	\sigma_{\rm bu}&=&\frac{2\pi}{2j+1}\sum_{m}\int_0^\infty \Bigg(\mel{\phi_{n_rljm}}{\,\big|S_{cT}S_{{\rm n}T}\big|^2\,}{\phi_{n_rljm}} \nonumber\\
& &\left.-\sum_{m'}\left|\mel{\phi_{n_rljm'}}{S_{cT}S_{{\rm n}T}}{\phi_{n_rljm}}\right|^2\right)b\,db.\label{eq3}
\end{eqnarray}
		The eikonal $S$-matrices in Eqs.~\eq{eq5} and \eq{eq3} are obtained  from integrating the action of the corresponding optical potential along the coordinate $Z$
	\begin{equation}
	S_{{(\scriptstyle {\rm n}, c)}{\scriptstyle T}{\scriptscriptstyle }}=\exp\left[-\frac{i}{\hbar v}\int_{-\infty}^{+\infty} U_{{(\scriptstyle {\rm n}, c)}{\scriptstyle T}{\scriptscriptstyle }} \left( R_{{(\scriptstyle {\rm n}, c)}{\scriptstyle T}{\scriptscriptstyle }}\right)\,dZ\right] \label{eq4}
	\end{equation} 
	with $R_{{(\scriptstyle {\rm n}, c)}{\scriptstyle T}{\scriptscriptstyle }}=\sqrt{b_{{(\scriptstyle {\rm n}, c)}{\scriptstyle T}{\scriptscriptstyle }}^2+Z^2}$ the n-$T$ and $c$-$T$ distances and $v$ the  projectile initial velocity.

\begin{figure}
	\centering 
		\includegraphics[width=\linewidth]{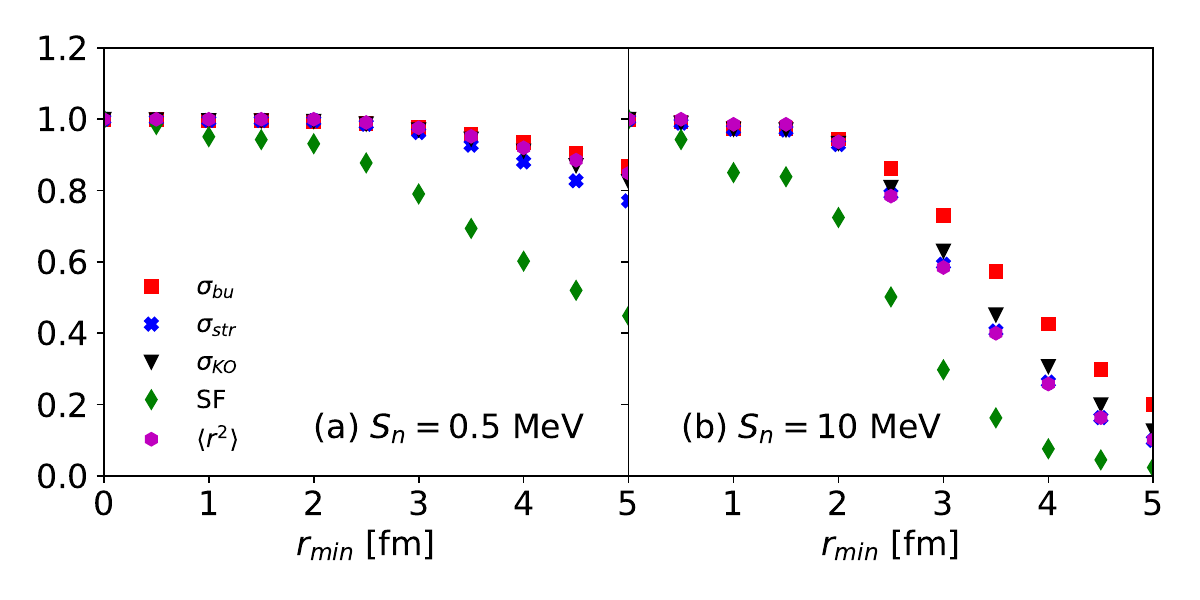}

	\caption{Analysis of the sensitivity of knockout cross sections $\sigma_{\rm KO}$, its diffractive-breakup $\sigma_{\rm bu}
	$ and stripping  $\sigma_{\rm str}$  contributions to the short range of the overlap wave function. 
	These reaction observables are computed for different cutoffs $r_{\rm min}$, below which the overlap function is put to 0; for readability they are normalised to the actual value obtained  with $r_{\rm min}=0$.
	The effect of the cutoff on the spectroscopic factor $SF$ and mean square radius $\langle r^2\rangle$ are shown as well.
	(a) loosely bound neutron ($S_{\rm n}=0.5$ MeV); (b) deeply bound neutron ($S_{\rm n}=$10 MeV).\label{Fig1}}	
\end{figure}

To investigate the sensitivity of knockout observables on the short-range part of the projectile wave function, we compute the cross sections \eq{eq5} and \eq{eq3} with a cutoff $r_{\rm min}$, below which the radial part of the wave function $\phi_{n_rljm}$ is set to zero.
The ratio of these results to the actual cross sections, {viz.} obtained with $r_{\rm min}=0$, are plotted as function of the cutoff in \Fig{Fig1}~(a) for the loosely-bound state ($S_{\rm n}=0.5$~MeV) and (b) for the deeply-bound state ($S_{\rm n}=10$~MeV).
The diffractive breakup component $\sigma_{\rm bu}$ is shown by the red squares, the stripping component $\sigma_{\rm str}$ by the blue crosses and their sum $\sigma_{\rm KO}$ by the black triangles.
As already seen in Refs.~\cite{HC19,HC21ANC}, in the one-neutron halo case, both contributions to the KO cross section are insensitive to the internal part of the overlap wave function.
Up to $r_{\rm min}\approx2.5$~fm, the cross sections are the same whether we set the overlap wave function to 0 or not [Fig.~\ref{Fig1}~(a)].
For the deeply-bound projectile [Fig.~\ref{Fig1}~(b)], that value is reduced to $r_{\rm min}\approx1.5$~fm because when $S_{\rm n}$ increases, the spatial extension of the wave function is reduced. This result hence confirms the peripherality of these KO observables.
At larger $r_{\rm min}$, we observe that, {for both $S_{\rm n}$,} $\sigma_{\rm bu}$ decays more slowly than $\sigma_{\rm str}$, showing that this observable is the most peripheral of the two.
The total KO cross section follows its dominant contribution: the diffractive breakup in the loosely bound case and the stripping in the deeply bound case.

To grasp how this result on the reaction observables relates to the structure of the projectile, we display in \Fig{Fig1} the evolution of  the spectroscopic factor $S_{1s1/2}$ (green diamonds) and the mean square radius $\langle r^2 \rangle$ (magenta hexagons) with $r_{\rm min}$, viz.\ we plot the ratio of either of these quantities obtained when the radial wave function is set to zero below $r_{\rm min}$ to their actual value.
Contrary to the KO cross sections, the spectroscopic factor is significantly sensitive to the short-range part of the overlap wave function,  and  therefore decays much faster than any of the cross sections with $r_{\rm min}$.
Setting the s.p.\ wave function to zero, even below 0.5~fm, leads to a visible reduction of the SF, whereas it does not affect either of the cross sections.
On the contrary, the mean square radius follows the $r_{\rm min}$ dependence of the KO observables; more precisely, it follows closely the dominant contribution to $\sigma_{\rm KO}$.

These results are very general: they do not depend on the neutron separation energy (we observe similar behavior for $S_{\rm n}\sim 0.5$--20~MeV), the geometry of the optical potentials, the number of nodes and the angular momentum of the overlap function. Although applying a cutoff in $r$ to  the s.p.\ wave functions is unrealistic, the insensitivity of the knockout cross sections to the short-range distance suggests that spectroscopic factors extracted from knockout data suffer from sizeable model uncertainties associated with the choice of the {s.p.} functions.
However,  knockout reaction provide an ideal probe to the mean square radius of the s.p.\ overlap function.

This simple linear dependence between the knockout cross sections and the  mean square radius of the overlap function $\langle r^2\rangle$ can be understood analytically from the expressions of their contributions \eq{eq5} and \eq{eq3} at {leading order}.
For projectiles with large core-to-neutron mass ratio, $\ve{R}_{cT}\sim \ve{R}$. 
The nuclear part of the  $c$-$T$ interaction thus plays a negligible role in the dynamics and is mainly inducing absorption at small $R$, i.e., at small impact parameters $b$ \cite{SAJ02,HC21ERT}.
Because knockout reactions are measured at intermediate energy,  $U_{{\rm n}T}/(\hbar v)$ remains small and we can approach the eikonal phase $S_{{\rm n}T}$ \eq{eq4} by the first terms of its Taylor expansion.
Using {a second}-order approximation of $\ve{R}_{{\rm n}T}$ in $\ve{R}$ and $\ve{r}$, the stripping cross section becomes	
\begin{eqnarray}
\sigma_{\rm str}&\approx&-\frac{4\pi}{\hbar v}\int_0^{\infty} |S_{cT}(b)|^2\left\{\int^{\infty}_{-\infty} W_{{\rm n}T}(R)\ dZ \right.\nonumber \\
	&& \left.+\frac{1}{3}\langle r^2\rangle\int^{\infty}_{-\infty}\left [\frac{1}{R}  W'_{{\rm n}T} (R)+\frac{1}{2}  W_{nT}''(R) \right] dZ\right\} b\,d b \label{eq5b},
	\end{eqnarray}	
where $W_{{\rm n}T}=\Im\left\{U_{{\rm n}T}\right\}$ and $W'_{{\rm n}T}$ and $W''_{{\rm n}T}$ are its first- and second-order derivatives, respectively.
The first term of this perturbative expression corresponds to a zero-range estimate of the stripping cross section.
The following two terms explain the direct relationship between the stripping component of the KO cross section and the mean square radius of the s.p.\ overlap functions observed in \Fig{Fig1}.

For the diffractive-breakup contribution, $S_{{\rm n}T}$ has to be expanded to the second order to obtain the non-vanishing expression
\begin{equation}
	\sigma_{\rm bu}\approx\frac{2\pi}{3\hbar^2v^2}\langle r^2\rangle\int^{+\infty}_0 |S_{cT}(b)|^2\left\|\int_{-\infty}^{+\infty} {\frac{1}{R}}U'_{nT}(R){\ve{R}}\ dZ \right\|^2 b\,d b. \label{eq7}
\end{equation}
This approximation explains the proportionality between $\sigma_{\rm bu}$ and $\langle r^2\rangle$ in \Fig{Fig1}.

\begin{figure}
	\centering
	\includegraphics[width=\linewidth]{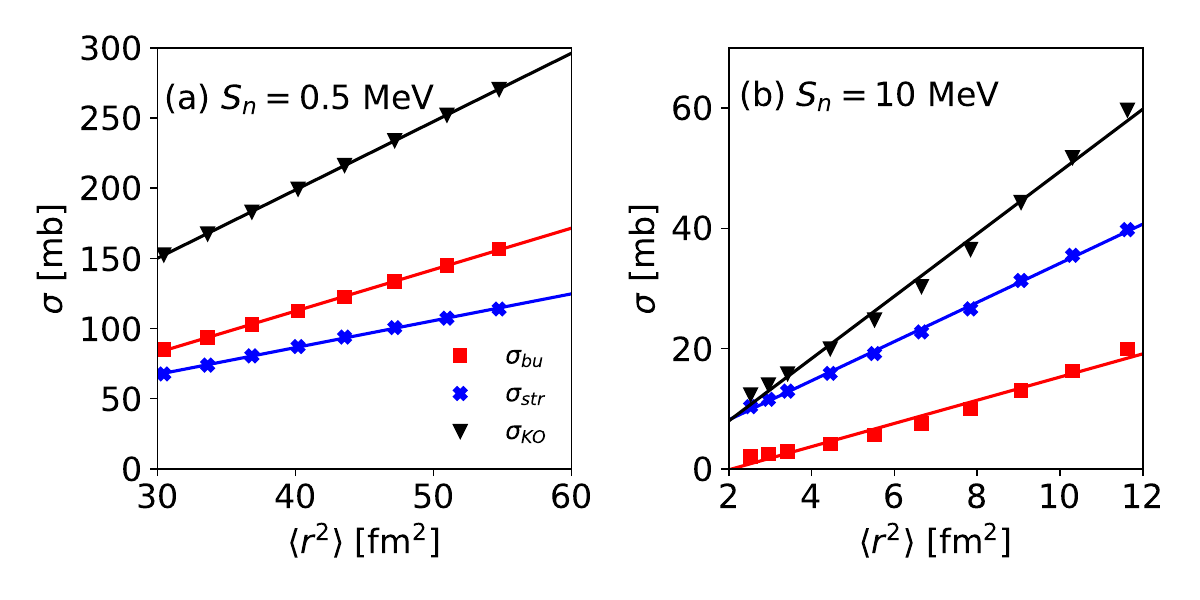} 
	\caption{Linear dependence of one-neutron knockout cross section $\sigma_{\rm KO}$, its diffractive-breakup $\sigma_{\rm bu}$ and its stripping $\sigma_{\rm str}$ contributions with the mean square radius of the s.p.\ state $\langle r^2\rangle$. (a) $S_{\rm n}=0.5$ MeV and (b) $S_{\rm n}=10$ MeV. The symbols are  numerical calculations and the lines are best least-square linear regressions to check the linearity.  \label{Fig2}}
\end{figure}

To verify the linear dependence of Eqs.~\eq{eq5b} and \eq{eq7}, we consider {a series of} s.p.\ wave functions {normalised to unity} with different $\langle r^2\rangle$ generated using different ranges of the Gaussian s.p.\ potential.
Figure~\ref{Fig2} shows KO cross sections (black triangles) and its two contributions $\sigma_{\rm bu}$ (red squares) and $\sigma_{\rm str}$ (blue crosses) as a function of  $\langle r^2\rangle$ for (a) $S_{\rm n}=0.5$ MeV and (b)~$S_{\rm n}=10$~MeV.
The linear dependence of knockout cross sections on the  $\langle r^2\rangle$ is clearly visible in both cases, as already noted  in Refs.~\cite{Gade:08,ABST17}. 

In the case of a halo nucleus projectile [Fig.~\ref{Fig2}~(a)], the cross sections scale perfectly with $\langle r^2\rangle$.   {Because} the neutron has a high probability of presence away from the core, the value of $\langle r^2\rangle$ {is directly} proportional to the ANC$^2$.
This is therefore in line  with our  previous analysis \cite{HC19}, which showed that, for one-neutron halo nuclei, $\sigma_{\rm bu}$ and $\sigma_{\rm str}$ scale with the ANC$^2$. 
{The fact that the ANC$^2$ fixes the value of} $\langle r^2\rangle$, and hence of $\sigma_{\rm KO}$, can therefore be {interpreted} as a signature of the halo structure.

For the removal of a deeply-bound neutron [Fig.~\ref{Fig2}(b)], we observe small deviations from the linear behavior.
{This non-linearity is most likely due to higher-order effects, which are expected to play a more significant role for short-ranged s.p.\  wave functions.
In particular,} the nuclear part of the  $c$-$T$ interaction has more influence on the dynamics of the reaction.
The factorization of $S_{cT}$ outside the matrix elements in Eqs.~\eq{eq5b} and \eq{eq7} is therefore less accurate~\cite{Hetal22}.

{In this Letter, we show that knockout cross sections are insensitive to the short-range part of the s.p.\ wave function but scale very accurately with the mean square radius of the overlap function.
We demonstrate analytically this dependence using a perturbation expansion of the eikonal  cross sections.
We have verified numerically that the linear relation in $\langle r^2\rangle$ is independent of the projectile binding energy, spatial extension,  spin and parity as well as of the geometry of the optical potentials simulating the interaction between the projectile components and the target.}

This work has important consequences since it suggests that various s.p.\ wave functions can reproduce the same knockout {cross section with different spectroscopic factors.
Accordingly,} spectroscopic factors extracted from such data suffer from sizeable model uncertainties.
{The analysis of KO data should account for these uncertainties.
In particular, they should include a careful study of the sensitivity of the nuclear-structure information inferred from the data to the geometry of the single-particle potentials.
This could be done, e.g., using a Bayesian approach as in Ref.~\cite{CLN23} for transfer reactions.
}

{Ideally, s.p.\ wave functions used in the analysis of KO cross sections should be constrained by other reaction observables.
In particular, the parallel-momentum distribution of the core, usually measured in KO experiments, could provide such additional constraint \cite{Gade:08}.
Using data from other reactions could also reduce the model uncertainty.
This is the spirit of our previous analyses performed on halo nuclei \cite{MYC19,HC21ANC}, in which it was shown that one description of $^{15}$C, constrained from transfer data, can also accurately describe breakup and KO cross sections.}

The present sensitivity analysis, and in particular the scaling of knockout cross sections with the mean square radius,  also sets up a clear connection between microscopic structure calculations and the few-body model of the reaction.
An exciting prospect is to extract neutron radii from one-neutron knockout data on neutron-rich nuclei.
{This would enable us to infer neutron-skin thicknesses, which can be translated into constraints on neutron-star radii \cite{ABST17,Thieletal19}.
{However, in that case too, the uncertainties due to the reaction model, i.e., the eikonal approximation, the choice of the optical potentials~\cite{OpticalPotentials} and of the s.p. function, need to be thoroughly quantified (see first studies in Refs.~\cite{HNLPRL22,Hetal22}). }
Such an alternative method to probe the neutron skin of nuclei could be valuable} since some usual methods, such as the coherent one-pion photoproduction, seem to be marred with significant uncertainty \cite{Col22}.

{The approximation of overlap functions by simple s.p.\ wave functions unavoidably introduces model uncertainties in reaction calculations.
This undoubtly affects the spectroscopic factors inferred from other reaction probes, such as transfer and quasifree scattering.
An analysis, similar to this one, should be performed to estimate the sensitivity of the cross sections of these reactions to the s.p. description of the projectile.
}

\paragraph{Acknowledgements}
The authors thank D. Bazin, F. M. Nunes and D. R. Phillips for constructive comments.
{C.~H. acknowledges the support  of the U.S. Department of Energy, Office of Science, Office of Nuclear Physics, under the FRIB Theory Alliance award no. DE-SC0013617 {and  under Work Proposal no. SCW0498}.} This work was prepared in part by LLNL under Contract no. DE-AC52-07NA27344.
This project has received funding from he Fonds de la 
Recherche Scientifique - FNRS under Grant Number 4.45.10.08,  the Deutsche Forschungsgemeinschaft Projekt-ID 279384907 -- SFB 1245, and the PRISMA+ (Precision Physics, Fundamental Interactions and Structure of Matter) Cluster of Excellence. 
P.~C. acknowledges the support of the State of Rhineland-Palatinate. This research was supported in part by the National Science Foundation under Grant No. NSF PHY-1748958.

\bibliography{HC_PRCKOSensitivityDeeplyBound}

\begin{thebibliography}{10}
\expandafter\ifx\csname url\endcsname\relax
  \def\url#1{\texttt{#1}}\fi
\expandafter\ifx\csname urlprefix\endcsname\relax\def\urlprefix{URL }\fi
\expandafter\ifx\csname href\endcsname\relax
  \def\href#1#2{#2} \def\path#1{#1}\fi

\bibitem{Oetal20}
T.~Otsuka, A.~Gade, O.~Sorlin, T.~Suzuki, Y.~Utsuno, Evolution of shell
  structure in exotic nuclei, Rev. Mod. Phys. 92 (2020) 015002.

\bibitem{HS01}
P.~G. Hansen, B.~M. Sherrill, Reactions and single-particle structure of nuclei
  near the drip lines, Nucl. Phys. A 693 (2001) 133--168, radioactive Nuclear
  Beams.

\bibitem{HT03}
P.~G. Hansen, J.~A. Tostevin, Direct reactions with exotic nuclei, Ann. Rev.
  Nucl. Part. Sc. 53 (2003) 219.

\bibitem{GG08}
A.~Gade, T.~Glasmacher, In-beam nuclear spectroscopy of bound states with fast
  exotic ion beams, Prog. Part. Nucl. Phys. 60 (2008) 161--224.

\bibitem{HM85}
M.~S. Hussein, K.~W. McVoy, Inclusive projectile fragmentation in the spectator
  model, Nucl. Phys. A 445 (1985) 124.

\bibitem{G59}
R.~J. Glauber, High energy collision theory, in: W.~E. Brittin, L.~G. Dunham
  (Eds.), Lecture in Theoretical Physics, Vol.~1, Interscience, New York, 1959,
  p. 315.

\bibitem{GM18}
M.~Gómez-Ramos, A.~M. Moro, {Binding-energy independence of reduced
  spectroscopic strengths derived from $(p,2p)$ and $(p,pn)$ reactions with
  nitrogen and oxygen isotopes}, Phys. Lett. B 785 (2018) 511.

\bibitem{PhysRevLett.104.112701}
J.~Lee, M.~B. Tsang, D.~Bazin, D.~Coupland, V.~Henzl, D.~Henzlova, M.~Kilburn,
  W.~G. Lynch, A.~M. Rogers, A.~Sanetullaev, A.~Signoracci, Z.~Y. Sun,
  M.~Youngs, K.~Y. Chae, R.~J. Charity, H.~K. Cheung, M.~Famiano, S.~Hudan,
  P.~O'Malley, W.~A. Peters, K.~Schmitt, D.~Shapira, L.~G. Sobotka,
  Neutron-proton asymmetry dependence of spectroscopic factors in ar isotopes,
  Phys. Rev. Lett. 104 (2010) 112701.

\bibitem{PhysRevLett.129.152501}
B.~P. Kay, T.~L. Tang, I.~A. Tolstukhin, G.~B. Roderick, A.~J. Mitchell,
  Y.~Ayyad, S.~A. Bennett, J.~Chen, K.~A. Chipps, H.~L. Crawford, S.~J.
  Freeman, K.~Garrett, M.~D. Gott, M.~R. Hall, C.~R. Hoffman, H.~Jayatissa,
  A.~O. Macchiavelli, P.~T. MacGregor, D.~K. Sharp, G.~L. Wilson, Quenching of
  single-particle strength in $a=15$ nuclei, Phys. Rev. Lett. 129 (2022)
  152501.

\bibitem{HNLPRL22}
C.~Hebborn, F.~M. Nunes, A.~E. Lovell, New perspectives on spectroscopic factor
  quenching from reactions, Phys. Rev. Lett. in print (2023).

\bibitem{Gade:08}
A.~Gade, P.~Adrich, D.~Bazin, M.~D. Bowen, B.~A. Brown, C.~M. Campbell, J.~M.
  Cook, T.~Glasmacher, P.~G. Hansen, K.~Hosier, S.~McDaniel, D.~McGlinchery,
  A.~Obertelli, K.~Siwek, L.~A. Riley, J.~A. Tostevin, D.~Weisshaar, Reduction
  of spectroscopic strength: Weakly-bound and strongly-bound single-particle
  states studied using one-nucleon knockout reactions, Phys. Rev. C 77 (2008)
  044306.

\bibitem{TG21}
J.~A. Tostevin, A.~Gade, Updated systematics of intermediate-energy
  single-nucleon removal cross sections, Phys. Rev. C 103 (2021) 054610.

\bibitem{Aetal21}
T.~Aumann, C.~Barbieri, D.~Bazin, C.~Bertulani, A.~Bonaccorso, W.~Dickhoff,
  A.~Gade, M.~Gómez-Ramos, B.~Kay, A.~Moro, T.~Nakamura, A.~Obertelli,
  K.~Ogata, S.~Paschalis, T.~Uesaka, Quenching of single-particle strength from
  direct reactions with stable and rare-isotope beams, Prog. Part. Nucl. Phys.
  118 (2021) 103847.

\bibitem{Fetal12}
F.~Flavigny, A.~Obertelli, A.~Bonaccorso, G.~F. Grinyer, C.~Louchart,
  L.~Nalpas, A.~Signoracci, Nonsudden limits of heavy-ion induced knockout
  reactions, Phys. Rev. Lett. 108 (2012) 252501.

\bibitem{Letal11}
C.~Louchart, A.~Obertelli, A.~Boudard, F.~Flavigny, Nucleon removal from
  unstable nuclei investigated via intranuclear cascade, Phys. Rev. C 83 (2011)
  011601.

\bibitem{GOMEZRAMOS2022137252}
M.~Gómez-Ramos, J.~Gómez-Camacho, A.~M. Moro, Binding-energy asymmetry in
  absorption explored through cdcc extended for complex potentials, Phys. Lett.
  B 832 (2022) 137252.

\bibitem{HP2022}
C.~Hebborn, G.~Potel, Green's function knockout formalism, Phys. Rev. C 107
  (2023) 014607.

\bibitem{PhysRevLett.103.202502}
C.~Barbieri, Role of long-range correlations in the quenching of spectroscopic
  factors, Phys. Rev. Lett. 103 (2009) 202502.

\bibitem{Jetal11}
O.~Jensen, G.~Hagen, M.~Hjorth-Jensen, B.~A. Brown, A.~Gade, Quenching of
  spectroscopic factors for proton removal in oxygen isotopes, Phys. Rev. Lett.
  107 (2011) 032501.

\bibitem{ATKINSON2019135027}
M.~Atkinson, W.~Dickhoff, Investigating the link between proton reaction cross
  sections and the quenching of proton spectroscopic factors in 48ca, Physics
  Letters B 798 (2019) 135027.

\bibitem{Wetal21}
J.~Wylie, J.~Oko\l{}owicz, W.~Nazarewicz, M.~P\l{}oszajczak, S.~M. Wang,
  X.~Mao, N.~Michel, Spectroscopic factors in dripline nuclei, Phys. Rev. C 104
  (2021) L061301.

\bibitem{HC19}
C.~Hebborn, P.~Capel, Sensitivity of one-neutron knockout to the nuclear
  structure of halo nuclei, Phys. Rev. C 100 (2019) 054607.

\bibitem{HC21ANC}
C.~Hebborn, P.~Capel, Halo effective field theory analysis of one-neutron
  knockout reactions of $^{11}\mathrm{Be}$ and $^{15}\mathrm{C}$, Phys. Rev. C
  104 (2021) 024616.

\bibitem{BC12}
D.~Baye, P.~Capel, Breakup reaction models for two- and three-cluster
  projectiles, in: C.~Beck (Ed.), Clusters in Nuclei, Vol. 2, Vol. 848,
  Springer, Heidelberg, 2012, p. 121.

\bibitem{HBE96}
K.~Hencken, G.~Bertsch, H.~Esbensen, Breakup reactions of the halo nuclei
  $^{11}\mathrm{Be}$ and $^{8}\mathrm{B}$, Phys. Rev. C 54 (1996) 3043.

\bibitem{SAJ02}
N.~C. Summers, J.~S. Al-Khalili, R.~C. Johnson, Nonadiabatic corrections to
  elastic scattering of halo nuclei, Phys. Rev. C 66 (2002) 014614.

\bibitem{HC21ERT}
C.~Hebborn, P.~Capel, Detailed study of the eikonal reaction theory for the
  breakup of one-neutron halo nuclei, Phys. Rev. C 103 (2021) 064614.

\bibitem{ABST17}
T.~Aumann, C.~A. Bertulani, F.~Schindler, S.~Typel, Peeling off neutron skins
  from neutron-rich nuclei: Constraints on the symmetry energy from
  neutron-removal cross sections, Phys. Rev. Lett. 119 (2017) 262501.

\bibitem{Hetal22}
C.~Hebborn, T.~R. Whitehead, A.~E. Lovell, F.~M. Nunes, Quantifying
  uncertainties due to optical potentials in one-neutron knockout reactions,
  Phys. Rev. C 108 (2023) 014601.

\bibitem{CLN23}
M.~Catacora-Rios, A.~E. Lovell, F.~M. Nunes, Complete quantification of
  parametric uncertainties in $(d,p)$ transfer reactions, Phys. Rev. C 108
  (2023) 024601.

\bibitem{MYC19}
L.~Moschini, J.~Yang, P.~Capel, $^{15}\mathrm{C}$: From halo effective field
  theory structure to the study of transfer, breakup, and radiative-capture
  reactions, Phys. Rev. C 100 (2019) 044615.

\bibitem{Thieletal19}
M.~Thiel, C.~Sfienti, J.~Piekarewicz, C.~J. Horowitz, M.~Vanderhaeghen, Neutron
  skins of atomic nuclei: per aspera ad astra, J. Phys. G 46 (2019) 093003.

\bibitem{OpticalPotentials}
C.~Hebborn, et~al., Optical potentials for the rare-isotope beam era, J. Phys.
  G: Nucl. Part. Phys. 50 (2023) 060501.

\bibitem{Col22}
F.~Colomer, P.~Capel, M.~Ferretti, J.~Piekarewicz, C.~Sfienti, M.~Thiel,
  V.~Tsaran, M.~Vanderhaeghen, Theoretical analysis of the extraction of
  neutron skin thickness from coherent $\ensuremath{\pi}{}^{0}$ photoproduction
  off nuclei, Phys. Rev. C 106 (2022) 044318.

\end{thebibliography}

\end{document}